\documentclass[prd,preprintnumbers,nofootinbib,superscriptaddress]{revtex4}
\usepackage[dvipdfmx]{graphicx}
\usepackage{enumerate}
\usepackage{color}
\usepackage{amsmath,amssymb}
\usepackage{mathrsfs}
\usepackage{bm}

\begin{document}
\preprint{CHIBA-EP-228v2, 2018.04.18}

\title{
Magnetic monopoles in pure $SU (2)$ Yang--Mills theory 
with a gauge-invariant mass
}

\author{Shogo Nishino}
\email{shogo.nishino@chiba-u.jp}
\affiliation{Department of Physics,  
Graduate School of Science, 
Chiba University, Chiba 263-8522, Japan
}

\author{Ryutaro Matsudo}
\email{afca3071@chiba-u.jp}
\affiliation{Department of Physics,  
Graduate School of Science and Engineering, 
Chiba University, Chiba 263-8522, Japan
}

\author{Matthias Warschinke}
\email{m_warschinke@chiba-u.jp}
\affiliation{Department of Physics,  
Graduate School of Science, 
Chiba University, Chiba 263-8522, Japan
}


\author{Kei-Ichi Kondo}
\email{kondok@faculty.chiba-u.jp}
\affiliation{Department of Physics,  
Graduate School of Science, 
Chiba University, Chiba 263-8522, Japan
}
\affiliation{Department of Physics,  
Graduate School of Science and Engineering, 
Chiba University, Chiba 263-8522, Japan
}

\begin{abstract}
In this paper, we show the existence of magnetic monopoles  in the pure $SU(2)$ Yang--Mills theory even in absence of scalar fields when the gauge-invariant mass term is introduced. 
This result follows from the recent proposal for obtaining the gauge field configurations in the Yang--Mills theory from the solutions of the field equations in the ``complementary'' gauge-scalar model with the radial length of the scalar field being fixed. 
The gauge-invariant mass term is obtained through a change of variables and a gauge-independent description of the Brout--Englert--Higgs mechanism which neither relies on the spontaneous breaking of gauge symmetry nor on the assumptions of the nonvanishing vacuum expectation value of the scalar field. 
According to these procedures, we solve under the static and spherically symmetric ansatz the field equations of the $SU(2)$ Yang--Mills theory coupled with an adjoint scalar field whose radial degree of freedom is fixed, and obtain a gauge field configuration of magnetic monopole with a minimum magnetic charge in the massive $SU(2)$ Yang--Mills theory. 
We compare the magnetic monopole obtained in this way in the massive Yang--Mills theory with the Wu--Yang magnetic monopole in the pure Yang--Mills theory and the 't Hooft--Polyakov magnetic monopole in the Georgi--Glashow model.


\end{abstract}

\maketitle

\section{Introduction}

In high energy physics, quark confinement is a long-standing problem to be solved in the framework of quantum chromo-dynamics (QCD). 
The dual superconductivity picture \cite{dual-superconductor} for the QCD vacuum is known as one of the most promising scenarios for quark confinement. 
For this hypothesis to be realized, however, condensations of some magnetic objects are indispensable.  
However, the relevant magnetic objects are not included in the action of QCD. 
Therefore, we begin our arguments with showing the existence of magnetic monopoles in the Yang--Mills theory even in absence of the scalar field.
Such magnetic monopoles in the pure Yang--Mills theory which we call \textit{Yang--Mills magnetic monopoles} should be compared with the 't Hooft--Polyakov magnetic monopole \cite{tH-P} in the Georgi--Glashow model which includes the scalar field in the action from the beginning, see, e.g., \cite{monopole, textbooks} for reviews of magnetic monopoles. 
The Yang--Mills magnetic monopoles are expected to be obtained as topological defects or topological solitons afterwards, since they are not included in the original QCD action. 


Nevertheless, we know \cite{textbooks} that the pure Yang--Mills theory with no scalar fields has the topological soliton only in the four-dimensional Euclidean space.  
Indeed, such topological solitons are known as instantons and antiinstantons, in agreement with the nontrivial homotopy group $\pi_{3} (S^{3}) = \mathbb{Z}$. 
This statement is no longer true once we introduce the mass term to the Yang--Mills theory, which we call the \textit{massive Yang--Mills theory} hereafter. 
In the massive Yang--Mills theory, it is shown that there exists the other topological soliton suggested from the nontrivial homotopy group $\pi_{2} (S^{2}) = \mathbb{Z}$. This is nothing but a magnetic monopole. 
It is reasonable to consider the massive Yang--Mills theory, in light of conjecture that the quantum Yang--Mills theory has a mass gap, even if the classical Yang--Mills theory is a conformal theory with no mass scale \cite{millenium}. 
Indeed, recent investigations arrive at a consensus that gluons behave as massive particles in the low-energy (momentum) region which is called the \textit{decoupling solution} \cite{decoupling}. 
In view of these, it is worth investigating the existence of magnetic monopole configurations in the massive Yang--Mills theory. 
However, a naive mass term for the gluon field breaks the gauge symmetry.


Recently, it has been shown that the gauge-invariant mass term  of the Yang--Mills field can be introduced by combining a change of variables and a gauge-independent description of the Brout--Englert--Higgs (BEH) mechanism \cite{Kondo2016} which neither relies on the spontaneous breaking of gauge symmetry nor on the assumptions of the nonvanishing vacuum expectation value of the scalar field.  
Moreover, it has been pointed out \cite{Kondo2016} that the Yang--Mills field configurations are obtained from the solutions of the field equations in the ``complementary'' gauge-scalar model with the scalar field  whose radial length is fixed
\footnote{ 
The ``complementarity'' originates from the confinement-Higgs complementarity in the gauge-scalar model which says that there is no phase transition between two phases, confinement and Higgs, which are analytically connected in the phase diagram \cite{complementarity}. 
See  \cite{Kondo2018} for the precise definition and more details for  ``complementarity''.
}.
According to these procedures, we show in this paper that  magnetic monopoles do exist in the pure $SU(2)$ Yang--Mills theory with a gauge-invariant mass term even in absence of scalar fields.
In order to obtain magnetic monopoles in the pure massive Yang--Mills theory, therefore, we can use the same procedures as those  used to obtain the 't Hooft--Polyakov magnetic monopoles in the Georgi--Glashow model.
In fact, we solve under the static and spherically symmetric ansatz the field equations of the $SU(2)$ Yang--Mills theory coupled to an adjoint scalar field whose radial degree of freedom is fixed.  Then we obtain a gauge field configuration for a magnetic monopole with a minimum magnetic charge in the massive $SU(2)$ Yang--Mills theory. 
We compare the magnetic monopole obtained in this way in the massive Yang--Mills theory with the Wu--Yang magnetic monopole \cite{Wu--Yang} in the pure Yang--Mills theory and the 't Hooft--Polyakov magnetic monopole in the Georgi--Glashow model.

The gauge-independent BEH mechanism \cite{Kondo2016} and the resulting separation $\mathscr{A}(x) = \mathscr{W}(x) + \mathscr{R}(x)$ of the gauge field $\mathscr{A}(x)$ into the massive mode $\mathscr{W}(x)$ and the residual mode $\mathscr{R}(x)$ in the gauge-scalar model provides a natural understanding of the gauge field decomposition in the Yang--Mills theory called the Cho--Duan--Ge--Faddeev--Niemi--Shabanov (CDGFNS) decomposition \cite{decomposition}, 
and the subsequent reformulations of the Yang--Mills theory using the new field variables \cite{Kondo-Murakami-Shinohara, Kondo-Murakami-Shinohara2006}, see e.g., \cite{Kondo-Kato-Shibata-Shinohara} for a review. 
In the CDGFNS decomposition the gauge field $\mathscr{A} (x) $ is decomposed into the two pieces: $\mathscr{A} (x) = \mathscr{V} (x) + \mathscr{X} (x)$, where $\mathscr{V} (x) $ is called the restricted (or residual) field which transforms in the same way as the original gauge field $\mathscr{A} (x) $ and  $\mathscr{X} (x) $ is called the remaining (or coset) field which transforms in the adjoint way under the gauge transformation $\mathscr{X} (x) \to U (x) \mathscr{X} (x) U^{-1} (x)$.  Consequently, we can
introduce a gauge invariant mass term  $M_{\mathscr{X}}^{2} \mathrm{tr} (\mathscr{X}_{\mu} \mathscr{X}^{\mu})$ in the Yang--Mills theory. 

The key ingredient in the CDGFNS decomposition is the so-called color direction field $\bm{n} (x)$ which transforms in the adjoint way under the local gauge transformation. 
However, the introduction of the color field prevents one from understanding the CDGFNS decomposition with ease. 
According to the gauge-independent BEH mechanism, the color field  $\bm{n}(x)$ in the reformulated Yang--Mills theory  is identified with the normalized adjoint scalar field $\hat{\bm{\phi}} (x)$ in the ``complementary'' gauge-scalar model. 
This way of introducing the color field will facilitate  understanding the role of the color field itself. 
The massive mode $\mathscr{W}(x)$ is identified with the remaining field $\mathscr{X}(x)$, while the residual mode $\mathscr{R}(x)$ with the restricted field $\mathscr{V}(x)$. 
Consequently, a gauge-invariant mass term  $M_{\mathscr{X}}^{2} \mathrm{tr} (\mathscr{X}_{\mu} \mathscr{X}^{\mu})$ in the reformulated Yang--Mills theory follows  according to the gauge-independent BEH mechanism from the kinetic term $(\mathscr{D}_{\mu} [\mathscr{A}] \bm{\phi}) \cdot ( \mathscr{D}^{\mu} [\mathscr{A}] \bm{\phi})=M_{\mathscr{W}}^{2} \mathrm{tr} (\mathscr{W}_{\mu} \mathscr{W}^{\mu})$ of the gauge-scalar model.


For this identification to work, we must solve an issue. 
The extended Yang--Mills theory written in terms of the field variables $( \mathscr{A}, \bm{n} )$ has extra degrees of freedom originated from the color field $\bm{n} (x)$ if we wish to obtain the gauge theory which is equipollent to the original Yang--Mills theory.
For this purpose, we impose the additional condition to relate the gauge field $\mathscr{A} (x)$ and the color field $\bm{n} (x)$ in such a way that the color field is given by a functional in terms of the gauge field: $\bm{n} = \bm{n} [\mathscr{A}]$. 
This condition is called the reduction condition. 
By using the resulting color field, we can define the magnetic  charge in a gauge-invariant way.
Such color field configurations are obtained from the gauge-scalar model, since it is shown \cite{Kondo2016} that the simultaneous solutions of the coupled field equations in the gauge-scalar model automatically satisfy the reduction condition.
Thus, we can construct gauge-invariant magnetic monopoles in the massive Yang--Mills theory using the color field obtained in this way.

It should be remarked that, within the framework of the reformulated Yang--Mills theory, the configurations of the color field $\bm{n} [\mathscr{A}]$  have been obtained by solving the reduction condition for a given configuration of the gauge field $\mathscr{A} (x)$, e.g., instantons and merons in \cite{Kondo-Fukui-Shibata-Shinohara}.
We now revisit this problem from the opposite direction such that the gauge field configurations are obtained for a given configuration of the color field or the normalized scalar field.


This paper is organized as follows.
In section II, we review how to obtain the gauge-invariant massive Yang--Mills theory (Yang--Mills theory with a gauge-invariant mass term) by starting from the ``complementary'' $SU (2)$ gauge-adjoint scalar model with a fixed radial degree of freedom.
In section III, we show by using the scaling argument due to Derrick \cite{Derrick} that there can exist  magnetic monopoles in the massive Yang--Mills theory.
In section IV, we obtain the magnetic monopole configuration with a minimum magnetic charge under the static and spherically symmetric ansatz. 
In section V, we discuss the short-distance and long-distance behavior of the gauge field and the chromo-magnetic field, in comparison with the 't Hooft--Polyakov magnetic monopole. 
For the gauge field, we also perform the decomposition based on the reformulation to investigate how the respective decomposed field behaves in the short-distance and long-distance regions.
In the final section, we discuss how the magnetic monopoles obtained in the massive Yang--Mills theory are responsible for quark confinement from the viewpoint of dual superconductivity and are consistent with the existence of a mass gap.  
In Appendix A, we summarize the essentials for the 't Hooft--Polyakov magnetic monopole. 
In Appendix B, we explain the method used for numerically solving the monopole equation.

\section{The massive Yang--Mills theory ``complementary'' to the gauge-adjoint scalar model}

In this section, we review the procedure \cite{Kondo2016} for obtaining the massive $SU(2)$ Yang--Mills theory from the ``complementary'' $SU (2)$ gauge-adjoint scalar model described by the Lagrangian density
\begin{equation}
\mathscr{L} = - \frac{1}{4} \mathscr{F}_{\mu \nu}^{A} \mathscr{F}^{\mu \nu A} + \frac{1}{2} \left( \mathscr{D}_{\mu}^{A B} [\mathscr{A}] \phi^{B} \right)  \left( \mathscr{D}^{\mu A C} [\mathscr{A}] \phi^{C} \right) + u \left( \phi^{A} \phi^{A} - v^{2} \right)
,
\label{Lagrangian}
\end{equation}
with
\begin{align}
\mathscr{F}_{\mu \nu}^{A} (x) = & \partial_{\mu} \mathscr{A}_{\nu}^{A} (x) - \partial_{\nu} \mathscr{A}_{\mu}^{A} (x) - g \epsilon^{A B C} \mathscr{A}^{B}_{\mu} (x) \mathscr{A}_{\nu}^{C} (x) , \\
\mathscr{D}_{\mu}^{A B} [\mathscr{A}] \phi^{B} (x) = & \partial_{\mu} \phi^{A} (x) - g \epsilon^{A B C} \mathscr{A}^{B}_{\mu} (x) \phi^{C} (x)
.
\label{eq_pre}
\end{align}
where $u = u (x)$ is the Lagrange multiplier field to incorporate the radially fixing constraint,
\begin{equation}
\phi^{A} (x)  \phi^{A} (x) = v^{2} 
, \ \ \ (A = 1,2,3, \ v > 0) .
\label{constraint}
\end{equation}
In what follows, we introduce respectively the inner and exterior products for the Lie-algebra-valued fields by
\begin{equation}
\mathcal{P} \cdot \mathcal{Q} := \mathcal{P}^{A} \mathcal{Q}^{A} , \ \ \ 
\mathcal{P} \times \mathcal{Q} := \epsilon^{A B C} T_{A} \mathcal{P}^{B} \mathcal{Q}^{C}
,
\end{equation}
with the generator of the Lie algebra $T_{A}$.

\begin{figure}[t]
\centering
\includegraphics[width=0.65\textwidth]{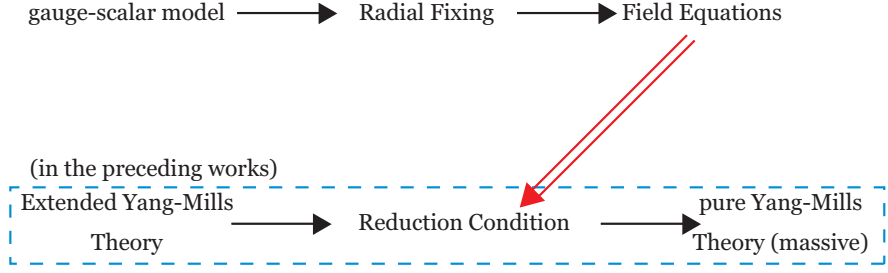}
\caption{The outline to obtain the massive Yang--Mills theory from the ``complementary'' gauge-scalar model.
The double-lined arrow stands for our approach in this paper.
The dotted box shows the approach in \cite{Kondo-Murakami-Shinohara2006,Kondo-Kato-Shibata-Shinohara}.
}
\label{equivalent}
\end{figure}

To begin with, we construct a composite vector boson  field $\mathscr{X}_{\mu} (x)$ from $\mathscr{A}_{\mu} (x)$ and $\hat{\bm{\phi}} (x)$ as
\begin{equation}
g \mathscr{X}_{\mu} (x) := \hat{\bm{\phi}} (x) \times \mathscr{D}_{\mu} [\mathscr{A}] \hat{\bm{\phi}} (x)
,
\end{equation}
where $\hat{\bm{\phi}} (x)$ is the normalized scalar field defined by
\begin{equation}
\hat{\bm{\phi}} (x) := \frac{1}{v} \bm{\phi} (x)
.
\end{equation}
Notice that $\mathscr{X}_{\mu} (x)$ transforms in the adjoint way under the gauge transformation $U (x) \in G =SU (2)$
\begin{align}
g \mathscr{X}_{\mu} (x) \to  
g \mathscr{X}_{\mu}^{\prime} (x)
=& \hat{\bm{\phi}}^{\prime} (x) \times \mathscr{D}_{\mu} [\mathscr{A}^{\prime}] \hat{\bm{\phi}}^{\prime} (x) 
= 
U (x) \hat{\bm{\phi}} (x) U^{\dagger} (x) \times U (x) \mathscr{D}_{\mu} [\mathscr{A}] U^{\dagger} (x) U (x) \hat{\bm{\phi}} (x) U^{\dagger} (x) 
\nonumber\\
= & U (x) \hat{\bm{\phi}} (x) \times \mathscr{D}_{\mu} [\mathscr{A}] \hat{\bm{\phi}} (x) U^{\dagger} (x)
= U (x) \mathscr{X}_{\mu} (x) U^{\dagger} (x)
.
\label{U-transf}
\end{align}
Then we find that the kinetic term $\mathscr{L}_{\rm kin} [ \mathscr{A} , \bm{\phi}]$ of the scalar field is identical to the mass term $\mathscr{L}_{\rm m}[\mathscr{X}]$ of the vector field $\mathscr{X}_{\mu} (x)$:
\begin{equation}
 \frac{1}{2} \mathscr{D}_{\mu} [\mathscr{A}] \bm{\phi} \cdot \mathscr{D}_{\mu} [\mathscr{A}] \bm{\phi} = \frac{1}{2} M_{\mathscr{X}}^{2} \mathscr{X}_{\mu} \cdot \mathscr{X}^{\mu} , \ \ \ 
M_{\mathscr{X}} : = g v
,
\end{equation}
as far as the radial degree of freedom of the scalar field is fixed \cite{Kondo2016}.
Thanks to (\ref{U-transf}), it is obvious that the obtained mass term of $\mathscr{X}_{\mu} (x)$ is gauge-invariant.
Therefore, $\mathscr{X}_{\mu} (x)$ can become massive without breaking the original gauge symmetry.
This gives a gauge-independent definition of the massive modes of the gauge field in the operator level.
It should be emphasized that we do not choose a specific vacuum with a non-vanishing vacuum expectation value $\langle 0| \phi(x) | 0 \rangle \not=0$ and hence need no spontaneous symmetry breaking.

By using the definition of the massive vector field $\mathscr{X}_{\mu} (x)$, the original gauge field $\mathscr{A}_{\mu} (x)$ is separated into two pieces:
\begin{equation}
\mathscr{A}_{\mu} (x) = \mathscr{X}_{\mu} (x) + \mathscr{V}_{\mu} (x)
,
\end{equation}
where the residual field $\mathscr{V}_{\mu} (x)$ can also be written in terms of $\mathscr{A}_{\mu} (x)$ and $\hat{\bm{\phi}} (x)$:
\begin{equation}
g \mathscr{V}_{\mu} (x) = g \mathscr{A}_{\mu} (x) - g \mathscr{X}_{\mu} (x) 
= g c_{\mu} (x) \hat{\bm{\phi}} (x) - \hat{\bm{\phi}} (x) \times \partial_{\mu} \hat{\bm{\phi}} (x)
, \ \ \ c_{\mu} (x) := \mathscr{A}_{\mu} (x) \cdot \hat{\bm{\phi}} (x)
.
\end{equation}

Next, we regard a new set of field variables $\{ c_{\mu} (x) , \mathscr{X}_{\mu} (x) , \hat{\bm{\phi}} (x) \}$ as obtained from the original field variables $\{ \mathscr{A}_{\mu} (x) , \hat{\bm{\phi}} (x) \}$ based on change of variables:
\begin{equation}
\{ \mathscr{A}_{\mu}^A (x) , \hat{\phi}^a (x) \} \to
\{ c_{\mu} (x) , \mathscr{X}_{\nu}^B (x) , \hat{\phi}^b (x) \} 
.
\label{cov}
\end{equation}
We wish to obtain the (pure) Yang--Mills theory from the ``complementary'' gauge-scalar model 
by identifying $c_{\mu} (x) , \mathscr{X}_{\mu} (x)$ and $\hat{\bm{\phi}} (x)$ with the fundamental field variables for describing the massive Yang--Mills theory anew, which means that we should perform the quantization in terms of  the field variables $\{ c_{\mu} (x) , \mathscr{X}_{\mu} (x) , \hat{\bm{\phi}} (x) \}$ appearing in the path-integral measure, as shown below.
The normalized scalar field $\hat{\bm{\phi}} (x)$ can be identified with the color field $\bm{n} (x)$ in the preceding approach \cite{Kondo-Murakami-Shinohara2006,Kondo-Kato-Shibata-Shinohara}.
  (See the dotted blue box in Fig.\ref{equivalent}.)

In the gauge-scalar model, $\mathscr{A}_{\mu} (x)$ and $\hat{\bm{\phi}} (x)$ are independent field variables. However, the pure Yang--Mills theory should be described by $\mathscr{A}_{\mu} (x)$ alone.
Hence the scalar field $\bm{\phi} (x)$ must be supplied by the gauge field $\mathscr{A}_{\mu} (x)$ in some way due to the strong interactions, or in other words, $\bm{\phi} (x)$ should be given as a functional of the gauge field $\mathscr{A}_{\mu} (x)$.
We take care of this discrepancy as follows. 
The independent degrees of freedom of the original gauge field $\mathscr{A}_{\mu}^{A} (x)$ in pure $SU (2)$ Yang--Mills theory are $[ \mathscr{A}_{\mu}^{A} (x) ] = 3 \times D = 3D$ per space-time point. Notice that we work in the off-shell counting. 
On the other hand, the new field variables have degrees of freedom: $[ \hat{\bm{\phi}} (x) ] = 2$,  $[ c_{\mu} (x) ] = D$, and $[ \mathscr{X}_{\mu}^{A} (x) ] = 2 \times D = 2D$, since  
the massive vector field $\mathscr{X}_{\mu} (x)$ obeys the condition,
\begin{equation}
\mathscr{X}_{\mu} (x) \cdot \hat{\bm{\phi}} (x) = 0
.
\end{equation}
We therefore observe that the theory with the new field variables has two extra degrees of freedom compared with the pure Yang--Mills theory. 
These extra degrees of freedom are eliminated by imposing the two constraints which we call the {\it reduction condition}.
We choose e.g.,
\begin{equation}
\bm{\chi} (x) := \hat{\bm{\phi}} (x) \times \mathscr{D}^{\mu} [\mathscr{A}] \mathscr{D}_{\mu} [\mathscr{A}] \hat{\bm{\phi}} (x) = 0
.
\label{reduction2}
\end{equation}
The reduction condition indeed eliminates the two extra degrees of freedom introduced by the radially fixed scalar field into the Yang--Mills theory, since it satisfies
\begin{equation}
\bm{\chi} (x) \cdot \hat{\bm{\phi}} (x) = 0
.
\end{equation}

The above consideration is incorporated into the path integral framework. 
But the following is just a sketch of the procedures, see sections 4.4, 4.5 and 5.6 of \cite{Kondo-Kato-Shibata-Shinohara} for precise statements.
Following the Faddeev--Popov procedure, we insert the unity to the functional integral to incorporate the reduction condition:
\begin{equation}
1 = \int \mathcal{D} \bm{\chi}^{\theta} \ \delta \left( \bm{\chi}^{\theta} \right) 
= \int \mathcal{D} \bm{\theta} \ \delta \left( \bm{\chi}^{\theta} \right) \Delta^{\rm red}
,
\end{equation}
where $\bm{\chi}^{\theta} := \bm{\chi} [ \mathscr{A} , \bm{\phi}^{\theta} ]$ is the reduction condition written in terms of $\mathscr{A}_{\mu} (x)$ and $\bm{\phi}^{\theta}$ which is the local rotation of $\bm{\phi} (x)$ by $\bm{\theta} = \bm{\theta} (x) = \theta^{A} (x) T_{A}$ and $\Delta^{\rm red} := \det \left( \frac{\delta \bm{\chi}^{\theta}}{\delta \bm{\theta}} \right)$ denotes the Faddeev--Popov determinant associated with the reduction condition $\bm{\chi} = 0$.
Then, we obtain after factoring out the gauge volume $\int \mathcal{D} \bm{\theta}$
\begin{align}
Z = & \int \mathcal{D} \hat{\bm{\phi}} \mathcal{D} \mathscr{A} \ \delta \left( \bm{\chi} \right) \Delta^{\rm red} \exp \left\{ i S_{\rm YM} [\mathscr{A}] + i S_{\rm kin} [ \mathscr{A} , \bm{\phi}] \right\} \nonumber\\
= & \int \mathcal{D} \hat{\bm{\phi}} \mathcal{D} c \mathcal{D} \mathscr{X} \ J \delta \left( \widetilde{\bm{\chi}} \right) \widetilde{\Delta}^{\rm red} \exp \left\{ i S_{\rm YM} [ \mathscr{V} + \mathscr{X}] + i S_{\rm m} [ \mathscr{X} ] \right\}
,
\label{path_integral}
\end{align}
where 
\begin{align}  
 S_{\rm kin} [ \mathscr{A} , \bm{\phi}] := \int d^Dx \frac{1}{2} \mathscr{D}_{\mu} [\mathscr{A}] \bm{\phi} \cdot \mathscr{D}_{\mu} [\mathscr{A}] \bm{\phi}, \quad
S_{\rm m}[\mathscr{X}] := \int d^Dx  \frac{1}{2} M_{\mathscr{X}}^{2} \mathscr{X}_{\mu} \cdot \mathscr{X}^{\mu} .
\end{align}
Here the reduction condition $\widetilde{\bm{\chi}}=0$ and the associated Faddeev-Popov determinant $\widetilde{\Delta}^{\rm red}$ are written in terms of the new variables. 
The Jacobian $J$ associated with the change of variables (\ref{cov}) is equal to one, $J = 1$ \cite{Kondo-Kato-Shibata-Shinohara}.
Therefore, we obtain the massive Yang--Mills theory which keeps the original gauge symmetry:
\begin{equation}
\mathscr{L}_{\rm mYM} = - \frac{1}{4} \mathscr{F}_{\mu \nu} [ \mathscr{V} + \mathscr{X} ] \cdot \mathscr{F}^{\mu \nu} [ \mathscr{V} + \mathscr{X} ] + \frac{1}{2} M_{\mathscr{X}}^{2} \mathscr{X}_{\mu} \cdot \mathscr{X}^{\mu}
, \ \ \ M_{\mathscr{X}} : = g v > 0 .
\label{mYM}
\end{equation}
We want to emphasize that the original gauge symmetry is kept at any stage in deriving the pure Yang-Mills theory from the gauge-scalar model and hence gauge symmetry is not broken before imposing the specific gauge fixing condition. Notice that the ordinary gauge-fixing part is omitted in the above discussion and in (\ref{path_integral}) for simplifying the notation.

\begin{figure}[t]
\centering
\includegraphics[width=0.35\textwidth]{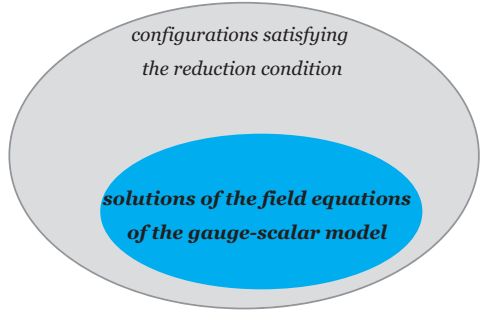}
\caption{The relation between the solutions of the field equations of the gauge-scalar model and the reduction condition.}
\label{inclusion}
\end{figure}

Finally, it should be remarked that the solutions of the classical field equations of the gauge-scalar model with a radial fixed scalar field  satisfy the reduction condition automatically.
This fact can be used to obtain the ``field configurations of path integral'' satisfying the reduction condition from the ``solutions of field equations'' of the corresponding gauge-scalar model. 
(But the converse is not true.) 
Fig.\ref{inclusion} shows the relation between the solutions of the field equations of the gauge-adjoint scalar model and the field configurations satisfying the reduction condition.
This fact is shown as follows. 
The field equations besides (\ref{constraint}) are obtained as
\begin{align}
\mathscr{D}^{\mu} [\mathscr{A}] \mathscr{F}_{\mu \nu} + g  \bm{\phi} \times \mathscr{D}_{\nu} [\mathscr{A}] \bm{\phi}  = & 0 , 
\label{eq_gauge}
\\
\mathscr{D}^{\mu} [\mathscr{A}] \mathscr{D}_{\mu} [\mathscr{A}] \bm{\phi}  - 2 u \bm{\phi} = & 0
.
\label{eq_scalar}
\end{align}
To eliminate the Lagrange multiplier field $u$ in (\ref{eq_scalar}) we take the inner product of (\ref{eq_scalar}) and $\bm{\phi} (x)$ and use (\ref{constraint}) to obtain
\begin{equation}
u = \frac{1}{2 v^{2}} \bm{\phi} \cdot \left( \mathscr{D}^{\mu} [\mathscr{A}] \mathscr{D}_{\mu} [\mathscr{A}] \bm{\phi} \right) = \frac{1}{2} \hat{\bm{\phi}} \cdot \left( \mathscr{D}^{\mu} [\mathscr{A}] \mathscr{D}_{\mu} [\mathscr{A}] \hat{\bm{\phi}} \right)
.
\end{equation}
The field equations (\ref{eq_gauge}) and (\ref{eq_scalar}) are rewritten in terms of $\mathscr{A}_{\mu} (x)$ and $\hat{\bm{\phi}} (x)$:
\begin{align}
\mathscr{D}^{\mu} [\mathscr{A}] \mathscr{F}_{\mu \nu} + g v^{2} \hat{\bm{\phi}} \times  \mathscr{D}_{\nu} [\mathscr{A}] \hat{\bm{\phi}} = & 0 , 
\label{eq_gauge1} \\
\mathscr{D}^{\mu} [\mathscr{A}] \mathscr{D}_{\mu} [\mathscr{A}] \hat{\bm{\phi}}  - \left( \hat{\bm{\phi}} \cdot  \mathscr{D}^{\mu} [\mathscr{A}] \mathscr{D}_{\mu} [\mathscr{A}] \hat{\bm{\phi}} \right) \hat{\bm{\phi}} = & 0
\label{eq_scalar1}
.
\end{align}
By applying the covariant derivative $\mathscr{D}^{\nu} [\mathscr{A}]$ to the equation (\ref{eq_gauge1}), the reduction condition is naturally induced:
\begin{align}
0 = & - \mathscr{D}^{\nu} [\mathscr{A}] \mathscr{D}^{\mu} [\mathscr{A}] \mathscr{F}_{\mu \nu} 
= g v^{2} \hat{\bm{\phi}} \times \mathscr{D}^{\nu} [\mathscr{A}] \mathscr{D}_{\nu} [\mathscr{A}] \hat{\bm{\phi}} = g v^{2} \bm{\chi}
.
\end{align}
Moreover,  by taking the exterior product of (\ref{eq_scalar1}) and $\hat{\bm{\phi}} (x)$, the reduction condition is induced again:
\begin{align}
0 = & \hat{\bm{\phi}} \times \mathscr{D}^{\mu} [\mathscr{A}] \mathscr{D}_{\mu} [\mathscr{A}] \hat{\bm{\phi}} - \left( \hat{\bm{\phi}} \cdot \mathscr{D}^{\mu} [\mathscr{A}] \mathscr{D}_{\mu} [\mathscr{A}] \hat{\bm{\phi}} \right) \left( \hat{\bm{\phi}} \times \hat{\bm{\phi}} \right) \nonumber\\
= & \hat{\bm{\phi}} \times \mathscr{D}^{\mu} [\mathscr{A}] \mathscr{D}_{\mu} [\mathscr{A}] \hat{\bm{\phi}} = \bm{\chi}
.
\end{align}
Hence, the simultaneous solutions of the coupled field equations (\ref{eq_gauge1}) and (\ref{eq_scalar1}) automatically satisfy the reduction condition (\ref{reduction2}).

The main purpose of this paper is to find the ``classical solutions'' of the coupled field equations (\ref{eq_gauge1}) and (\ref{eq_scalar1}) of the gauge-scalar model with a radial fixed scalar field  to obtain the ``field configurations'' satisfying the reduction condition (\ref{reduction2}) in the path-integral (\ref{path_integral}) of a massive quantum Yang--Mills theory.

\section{Scaling argument of the massive Yang--Mills theory}

In this section we examine the existence of the static and stable configuration in the massive Yang--Mills theory.
For this purpose, we follow the scaling argument due to Derrick \cite{Derrick}. 
Only in this section we consider an arbitrary spatial dimension $d$.

In the gauge-adjoint scalar model with a radial-fixing constraint, the static energy $E$ can be written after eliminating the Lagrange multiplier field $u (x)$ as
\begin{equation}
E = \int d^{d} x \biggl[ \frac{1}{4} \mathscr{F}_{j k} \cdot \mathscr{F}_{j k} + \frac{v^{2}}{2} \left( \mathscr{D}_{j} [\mathscr{A}] \hat{\bm{\phi}} \right) \cdot \left( \mathscr{D}_{j} [\mathscr{A}] \hat{\bm{\phi}} \right) + \frac{v^{2}}{2} \left( 1 - \hat{\bm{\phi}} \cdot \hat{\bm{\phi}}  \right)  \hat{\bm{\phi}} \cdot \mathscr{D}_{j} [\mathscr{A}] \mathscr{D}_{j} [\mathscr{A}] \hat{\bm{\phi}} \biggr]
.
\end{equation}
By rescaling the spatial variable $\bm{x}$ as $\bm{x} \to \mu \bm{x}$, the fields are transformed as $\Phi (\bm{x}) \to \Phi^{(\mu)} (\bm{x})$ in general: For the scalar and the vector field,  
\begin{align}
\hat{\bm{\phi}}^{(\mu)} (\bm{x} ) = \hat{\bm{\phi}} (\mu \bm{x}) , \ \ \ 
\mathscr{A}_{j}^{(\mu)} ( \bm{x} ) = \mu \mathscr{A}_{j} (\mu \bm{x}) 
,
\end{align}
which yields
\begin{align}
\left( \mathscr{D}_{j} [\mathscr{A}] \hat{\bm{\phi}} \right)^{(\mu)} (\bm{x}) = \mu \left( \mathscr{D}_{j} [\mathscr{A}] \hat{\bm{\phi}} \right) (\mu \bm{x}) , \ \ \ 
\mathscr{F}_{j k}^{(\mu)} (\bm{x}) = \mu^{2} \mathscr{F}_{j k} (\mu \bm{x})
.
\end{align}
Then  the scaled energy $E(\mu, d)$ obeys
\begin{equation}
E(\mu, d) = \mu^{4-d} E_{4} + \mu^{2-d} E_{2}
,
\label{scaling}
\end{equation}
where
\begin{align}
E_{4} := & \int d^{d} x \ \frac{1}{4} \mathscr{F}_{j k} \cdot \mathscr{F}_{j k} , \\
E_{2} := & \int d^{d} x \biggl[ \frac{v^{2}}{2} \left( \mathscr{D}_{j} [\mathscr{A}] \hat{\bm{\phi}} \right) \cdot \left( \mathscr{D}_{j} [\mathscr{A}] \hat{\bm{\phi}} \right) + \frac{v^{2}}{2} \left( 1 - \hat{\bm{\phi}} \cdot \hat{\bm{\phi}} \right) \hat{\bm{\phi}} \cdot \mathscr{D}_{j} [\mathscr{A}] \mathscr{D}_{j} [\mathscr{A}] \hat{\bm{\phi}} \biggr]
.
\end{align}

For the massive Yang--Mills theory (\ref{mYM}), the scaled energy $E(\mu , d)$ obeys the same equation as (\ref{scaling}) with the replacement:
\begin{equation}
E_{4} :=  \int d^{d} x \ \frac{1}{4} \mathscr{F}_{j k} \cdot \mathscr{F}_{j k} , \ \ \ \ 
E_{2} :=  \int d^{d} x \ \frac{1}{2} M_{\mathscr{X}}^{2} \mathscr{X}_{j} \cdot \mathscr{X}_{j}
.
\end{equation}
We find that $E(\mu, d)$ has a stationary point in $2 < d < 4$ spatial dimension:
\begin{equation}
\frac{d E(\mu , d)}{d \mu} = 0
\ \ \ \mathrm{at} \ \ \ 
\mu = \sqrt{\frac{(d-2) E_{2}}{(4-d) E_{4}}}
,
\end{equation}
implying that there can exist a stable configuration with a finite energy which differs from the vacuum configuration.
It should be noticed that such a stable configuration can exist {\it only} in $d =3$.
Therefore, we can obtain the static topological soliton in the $(3+1)$-dimensional massive Yang--Mills theory.
The explicit construction is given in the next section. 

This result should be compared with the pure (massless) Yang--Mills theory and the $SU(2)$ Georgi--Glashow model:
The scaled energies for these theories are respectively given by
\begin{align}
E_{\rm YM} (\mu, d) = & \mu^{4-d} E_{4} ,
\\
E_{\rm GG} (\mu, d) = & \mu^{4-d} E_{4} + \mu^{2-d} E_{2} + \mu^{-d} E_{0}   
,
\end{align}
where $E_{0}$ in $E_{\rm GG} (\mu, d)$ comes from the potential term (see Appendix~\ref{AppendixA}).
Notice that for the pure Yang--Mills theory only the term of the gauge field exists and hence there is no stationary point under the  scaling, which implies the non-existence of the static and stable soliton solutions in the $(3+1)$-dimensional massless Yang--Mills theory. 
For the Georgi--Glashow model, see Appendix~\ref{AppendixA} for the existence of the static soliton solution, i.e., the 't Hooft--Polyakov magnetic monopole.

\section{magnetic monopole configuration}

Because of the constraint (\ref{constraint}) the normalized scalar field $\hat{\bm{\phi}} (x)$ takes the value in the target space of the two-dimensional sphere $S^{2}$.
Then, by regarding $\hat{\bm{\phi}} (x)$ as the map
\begin{equation}
\hat{\bm{\phi}} (x) : S_{\rm phys}^{2} \to S_{\rm target}^{2} 
,
\end{equation}
there could exist the topological soliton solutions related to the nontrivial homotopy group $\pi_{2} ( S^{2} ) = \mathbb{Z}$.

For the gauge field $\mathscr{A}_{\mu}^{A} (x)$, we adopt the static and spherically symmetric ansatz of the 't Hooft--Polyakov type \cite{tH-P}:
\begin{equation}
g \mathscr{A}_{0}^{A} (x) = 0 , \ \ \ 
g \mathscr{A}_{j}^{A} (x) = \epsilon^{j A k} \frac{ x^{k}}{r } \frac{ 1 - f (r)}{r}
.
\end{equation}

For the scalar field $\bm{\phi} (x)$, we adopt the simplest static and spherically symmetric ansatz
\footnote{It should be noted that in this setup the reduction condition (\ref{reduction2}) is automatically satisfied due to its tensor structure (without knowing the profile functions):
\begin{align}
\bm{\chi} = & \hat{\bm{\phi}} \times \mathscr{D}^{\mu} [\mathscr{A}] \mathscr{D}_{\mu} [\mathscr{A}] \hat{\bm{\phi}} 
= \epsilon^{A B C} T_{A} \hat{\phi}^{B} \left( - \mathscr{D}_{j} [\mathscr{A}] \mathscr{D}_{j} [\mathscr{A}] \hat{\bm{\phi}} \right)^{C} \nonumber\\
= & - \epsilon^{A B C} T_{A} \frac{x^{B}}{r} \frac{x^{C}}{r} h (r) \biggl[ \frac{d^{2} h (r)}{d r^{2}} + \frac{2}{r} \frac{d h (r)}{d r} - \frac{2}{r^{2}} h (r) f^{2} (r)  \biggr] = 0
, \quad T_A=\frac{\sigma_A}{2} .
\end{align}}:
\begin{equation}
\phi^{A} (x) = v \frac{x^{A}}{r} h (r) \ \ \ \Longleftrightarrow \ \ \ 
\hat{\phi}^{A} (x) = \frac{x^{A}}{r} h (r)
.
\end{equation}
The profile functions $f (r)$ and $h (r)$ are unknown functions to be determined by solving the field equations.

In order to simplify our notations, we consider the Lagrangian $L$
\begin{equation}
L =  \int d^{3} x \ \mathscr{L} =  \int_{0}^{\infty} d r \  \widetilde{\mathscr{L}}
,
\end{equation}
to redefine the Lagrangian density by $\widetilde{\mathscr{L}} =  4 \pi r^{2} \mathscr{L}$:
\begin{equation}
\widetilde{\mathscr{L}} = \frac{4 \pi}{g^{2}} \biggl[- f^{\prime 2} (r) - \frac{( f^{2} (r) - 1)^{2}}{2 r^{2}} - \frac{1}{2} g^{2} v^{2} r^{2} h^{\prime 2} (r) - g^{2} v^{2} f^{2} (r) h^{2} (r) + u g^{2} v^{2} r^{2} \left( h^{2} (r) - 1 \right) \biggr]
,
\end{equation}
where the prime denotes a derivative with respect to $r$.

The equations for the profile functions $f (r)$ and $h (r)$ are obtained as
\begin{align}
f^{\prime \prime} (r) =&  \frac{f^{3} (r) - f (r)}{r^{2}} + g^{2} v^{2} h^{2} (r) f (r) ,
\label{eq_f} \\ 
\left( r^{2} h^{\prime}  (r) \right)^{\prime} = &  2 f^{2} (r) h (r) - 2 u r^{2} h (r) ,
\label{eq_h} \\ 
h^{2} (r) - 1=&  0
\label{eq_lambda}
.
\end{align}
Eq.(\ref{eq_lambda}) comes from the constraint and can be solved
\begin{equation}
h (r) = \pm 1
.
\label{ans_h}
\end{equation}
By substituting (\ref{ans_h}) into the other equations, we have
\begin{align}
f^{\prime \prime} (r) = & \frac{f^{3} (r) - f (r)}{r^{2}} + g^{2} v^{2} f (r) ,
\label{eq_f2} \\ 
0 = & f^{2} (r) - u r^{2} 
.
\end{align}
Thus we can determine the Lagrange multiplier field $u = u (r)$ by
\begin{equation}
u (r) = \frac{f^{2} (r)}{r^{2}}
,
\end{equation}
once the remaining equation (\ref{eq_f2}) which we call the monopole equation is solved.
By rescaling 
\begin{equation}
r \to \rho := M_{\mathscr{X}} r ,
\end{equation}
with $\rho$ now being dimensionless, the monopole equation is reduced to
\begin{equation}
f^{\prime \prime} (\rho) = \frac{f^{3} (\rho) - f (\rho)}{\rho^{2}} + f (\rho)
.
\label{eq_f3}
\end{equation}

First, we examine the asymptotic behavior of $f (r)$.
The static energy $E$ is given by
\begin{align}
E = & \frac{4 \pi}{g^{2}} \int_{0}^{\infty} d r \biggl[ f^{\prime 2} (r) + \frac{(f^{2} (r) - 1)^{2}}{2 r^{2}} + g^{2} v^{2} f^{2} (r) \biggr]  \nonumber\\
= & \frac{4 \pi M_{\mathscr{X}}}{g^{2}} \int_{0}^{\infty} d \rho \biggl[ f^{\prime 2} (\rho) + \frac{(f^{2} (\rho) - 1 )^{2}}{2 \rho^{2}} + f^{2} (\rho) \biggr]
, 
\end{align}
where in the second equality we have rescaled $r \to \rho$.
One can find the boundary conditions for $f (\rho)$ by requiring the energy $E$ to be finite:
\begin{equation}
f (\rho) \xrightarrow{\rho \to 0} \pm 1 + \mathcal{O} (\rho^{1/2} ) , \ \ \ \ \ 
f (\rho) \xrightarrow{\rho \to \infty} 0 + \mathcal{O} (\rho^{-1})
.
\label{bc}
\end{equation}
For small $\rho$, we further require 
\begin{equation}
f (\rho) \xrightarrow{\rho \to 0} +1 + \mathcal{O} (\rho^{\alpha}),  \ \alpha > 1,
\end{equation}
so that the gauge field $\mathscr{A}_{j}^{A} (x)$ becomes non-singular at the origin.

Here, one finds that $f (\rho)\equiv 0$ is a solution of the monopole equation (\ref{eq_f3}). 
This is nothing but the Wu--Yang magnetic monopole. 
However, this solution yielding $\mathscr{X}_{j} (x) = 0$ does not satisfy the boundary condition (\ref{bc}) for $\rho \approx 0$, which leads to infinite energy $E = \infty$.
Conversely, the solution $f (\rho) \neq 0$ means $\mathscr{X}_{\mu} (x) \neq 0$, which yields a finite energy $E < \infty$.

In order to obtain the asymptotic behavior of $f (\rho)$ for small $\rho$, let us define $f (\rho) = 1 + g (\rho)$ with $|g (\rho)| \ll 1$ and linearize the monopole equation (\ref{eq_f3}):
\begin{equation}
\rho^{2} g^{\prime \prime} (\rho) - 2 g (\rho)  - \rho^{2} g (\rho) = \rho^{2}
.
\label{eq_g}
\end{equation}
The linear differential equation (\ref{eq_g}) for $g (\rho)$ has the following general solution
\begin{align}
g (\rho) =& C_{1} \left( \cosh \rho - \frac{\sinh \rho}{\rho} \right) + C_{2}  \left( \frac{\cosh \rho}{\rho} - \sinh \rho \right) \nonumber\\
&  - 1 + \left( \cosh \rho - \frac{\sinh \rho}{\rho} \right) \mathrm{Chi} \ \rho + \left( \frac{\cosh \rho}{\rho} - \sinh \rho \right) \mathrm{Shi} \ \rho 
,
\label{sol_g1}
\end{align}
where we have introduced the hyperbolic cosine and sine integral  $\mathrm{Chi}\ x$ and $\mathrm{Shi}\ x$ respectively, defined with the Euler constant $\gamma$ by
\begin{equation}
\mathrm{Chi}\ x := \gamma + \log x + \int_{0}^{x} d t \  \frac{\cosh t - 1}{t} , \ \ \ 
\mathrm{Shi}\ x := \int_{0}^{x} d t \ \frac{\sinh t}{t}
.
\end{equation}
Here the first two terms of (\ref{sol_g1}) correspond to the general solution consisting of two independent special solutions $(\cosh \rho - \frac{\sinh \rho}{\rho})$ and $(\frac{\cosh \rho}{\rho} - \sinh \rho)$ of the homogeneous equation obtained by eliminating the inhomogeneous term $\rho^{2}$ of (\ref{eq_g}), and the remaining terms represent a special solution of the inhomogeneous equation (\ref{eq_g}).

Under the boundary conditions $g (0) = 0$ and $g^{\prime} (0) = 0$, we can determine only one coefficient $C_{2} = 0$:
\begin{equation}
g (\rho) = C_{1}  \left( \cosh \rho - \frac{\sinh \rho}{\rho} \right) - 1 + \left( \cosh \rho - \frac{\sinh \rho}{\rho} \right) \mathrm{Chi} \ \rho + \left( \frac{\cosh \rho}{\rho} - \sinh \rho \right) \mathrm{Shi} \ \rho
.
\label{sol_g}
\end{equation}


The Taylor expansion of the solution (\ref{sol_g}) around the origin $\rho = 0$ has the form
\begin{equation}
g (\rho) = \widetilde{C} \rho^{2} + \frac{1}{3} \rho^{2} \log \rho + \mathcal{O} (\rho^{4})
, \ \ \ \widetilde{C} := \frac{1}{9} \left( - 4 + 3 \gamma + 3 C_{1} \right)
.
\end{equation}
Thus, under the boundary conditions $f (0) = 1$ and $f^{\prime} (0) = 0$, we can set the asymptotic form of $f (\rho)$ around the origin:
\begin{equation}
f (\rho) \approx 1 + \widetilde{C} \rho^{2} + \frac{1}{3} \rho^{2} \log \rho + \cdots 
, \ \ \ \ (\rho \approx 0)
\end{equation}
where $\widetilde{C}$ is arbitrary at this stage.

For large $\rho$, we adopt the asymptotic form
\begin{equation}
f (\rho) \approx e^{- \rho} \sum_{n = 0}^{\infty} D_{n}  \rho^{- n}
. \ \ \ \ (r \approx \infty)
\end{equation}
In the similar way to the above, we can determine the coefficients $D_{n}$ as
\begin{equation}
f (\rho) \approx D_{0} e^{- \rho} \left( 1 - \frac{1}{2 \rho} + \frac{3}{8 \rho^{2}} - \cdots \right)
,
\end{equation}
where the overall factor $D_{0}$ is arbitrary at this stage.
The monopole equation (\ref{eq_f3}) can be solved in a numerical way, see Appendix \ref{AppendixB} for the detail.
The coefficients $\widetilde{C}$ and $D_{0}$ can be determined in a numerical way, as well.

\begin{figure}[t]
\centering
\includegraphics[width=0.45\textwidth]{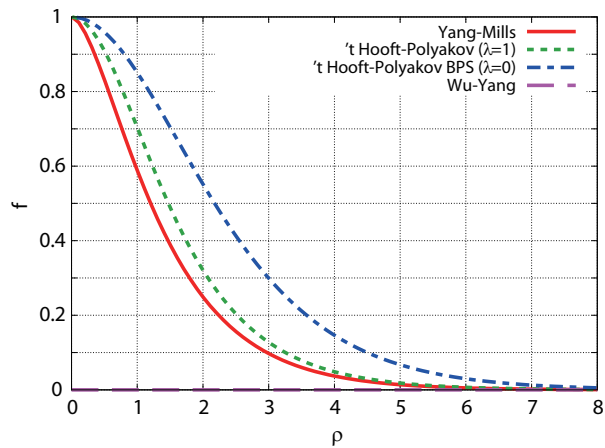}
\caption{The solution $f $ of the Yang--Mills monopole equation (\ref{eq_f3})  as a function of $\rho = M_{\mathscr{X}} r$  to be compared with the 't Hooft--Polyakov monopole solutions (for $\lambda =0$ and $\lambda=1$) and the Wu--Yang magnetic monopole with $f \equiv 0$.}
\label{solution}
\end{figure}

Fig.\ref{solution} shows the obtained solution $f (\rho)$  of the monopole equation (\ref{eq_f3}) as a function of $\rho$, which should be compared with the usual 't Hooft--Polyakov monopole solution. 

The obtained Yang--Mills magnetic monopole "configuration" is apparently similar to the infinitely large coupling limit $\lambda \to \infty$ of the standard 't Hooft-Polyakov magnetic monopole ``solution''. However, there is a crucial difference even in the classical level.  In the  't Hooft-Polyakov magnetic monopole, the scalar field is zero at the origin $\phi(0)=0$ for any value of $\lambda$, and this is the case even in the limit $\lambda \to \infty$ where $\phi(x) $ approaches $v$ for any $x$ except the origin $x=0$.  While, our magnetic monopole always guarantees $\phi(x)=v$ at every $x$ including the origin $\phi(0)=v$ due to the radially fixed constraint (\ref{eq_lambda}). 
In fact, we could not obtain the 't Hooft-Polyakov monopole in the limit $\lambda \to \infty$, since our numerical method cannot be applied to the large values of $\lambda$. But it's no concern of ours.

The details for the usual 't Hooft--Polyakov monopole in the Georgi--Glashow model are summarized in Appendix~\ref{AppendixA}.

From this numerical solution, we can calculate the static energy or the rest mass of a magnetic monopole $E$ as
\begin{equation}
E = \frac{4 \pi M_{\mathscr{X}}}{g^{2}} \int_{0}^{\infty} d \rho \biggl[ f^{\prime 2} (\rho) + \frac{(f^{2} (\rho) - 1)^{2}}{2 \rho^{2}} + f^{2} (\rho) \biggr] \approx \frac{4 \pi M_{\mathscr{X}}}{g^{2}} \times 1.78206
.
\end{equation}
This result also shows that the obtained solution $f (\rho)$ is different from the Bogomol'nyi--Prasad--Sommerfield (BPS) monopole \cite{BPS}:
By definition, the energy in the BPS limit is given by
\begin{equation}
E = \frac{4 \pi v}{g} = \frac{4 \pi M_{\mathscr{W}}}{g^{2}} , \ \ \ M_{\mathscr{W}} = g v
.
\end{equation}

We define the energy density $e (\rho)$ by
\begin{equation}
E = \int d^{3} x \ \mathscr{H} (r) = \int_{0}^{\infty} d \rho \ 4 \pi \rho^{2} \frac{M_{\mathscr{X}}}{g^{2}} \mathscr{H} (\rho) = \frac{M_{\mathscr{X}}}{g^{2}} \int_{0}^{\infty} d \rho \ e (\rho)
,
\end{equation}
where $\mathscr{H} (r)$ is the Hamiltonian density.
The energy density $e (\rho)$ can be written as
\begin{equation}
e (\rho) = 4 \pi \biggl[ f^{\prime 2} (\rho) + \frac{(f^{2} (\rho) - 1)^{2}}{2 \rho^{2}} + \frac{1}{2} \rho^{2} h^{\prime 2} (\rho) + f^{2} (\rho) h^{2} (\rho) + V (h^{2}) \biggr]
.
\label{energy_density}
\end{equation}

\begin{figure}[t]
\centering
\includegraphics[width=0.45\textwidth]{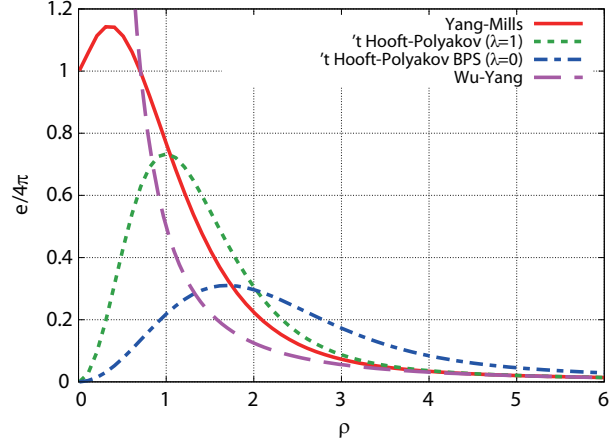} 
\caption{The energy density $e/ 4\pi$ of the Yang--Mills monopole as a function of $\rho = M_{\mathscr{X}} r$ to be compared with the 't Hooft--Polyakov monopoles (for $\lambda =0$ and $\lambda=1$) and the Wu--Yang magnetic monopole (diverging at the origin).}
\label{density}
\end{figure}

Fig.\ref{density} is the plot of the energy density $e (\rho)$ as a function of $\rho$ obtained from the solution $f (\rho)$, which should also be compared with the case of the 't Hooft--Polyakov monopoles. 
One can find that the energy density of the Yang--Mills monopole is much different from the 't Hooft--Polyakov solution at the origin.
This is caused by the radially fixed condition:
In the 't Hooft--Polyakov case, $e (0) = 0$ originates from $h (0) = 0$, which persists even in the limit $\lambda = \infty$, while in our case,  $h (0) = \pm 1$ with no potential term $V (h^2)= 0$, the contribution from the fourth term in (\ref{energy_density}) for $e (\rho)$ survives at the origin due to $f (0) = 1$.

\section{behaviors of gauge field and chromo-magnetic field}
\subsection{The Gauge Field}

We shall separate the gauge field $\mathscr{A}_{\mu} (x)$ into two pieces:
\begin{equation}
\mathscr{A}_{\mu} (x) = \mathscr{V}_{\mu} (x) + \mathscr{X}_{\mu} (x)
,
\end{equation}
where
\begin{equation}
g \mathscr{X}_{\mu} =  \hat{\bm{\phi}} \times  \mathscr{D}_{\mu} [\mathscr{A}] \hat{\bm{\phi}}  , \ \ \ 
g \mathscr{V}_{\mu} = g \mathscr{A}_{\mu} - g \mathscr{X}_{\mu}
= g ( \mathscr{A}_{\mu} \cdot \hat{\bm{\phi}} ) \hat{\bm{\phi}} +  \partial_{\mu} \hat{\bm{\phi}} \times \hat{\bm{\phi}}
.
\end{equation}
In the present ansatz, by using the normalized scalar field $\hat{\bm{\phi}} (x)$ with $h (r) = +1$ and the Pauli matrices $T_{A} = \frac{1}{2} \sigma_{A}$, 
\begin{equation}
\hat{\bm{\phi}} (x) =  \frac{x^{A}}{r} \frac{\sigma_{A}}{2}
, 
\end{equation}
they are explicitly written as
\begin{equation}
g \mathscr{V}_{j} (x) = \frac{\epsilon^{j A k} x^{k}}{r^{2}} \frac{\sigma_{A}}{2} , \ \ \ \ 
g \mathscr{X}_{j} (x) = - \frac{\epsilon^{j A k} x^{k}}{r^{2}} \frac{\sigma_{A}}{2} f (M_{\mathscr{X}} r)
, 
\end{equation}
and their time components vanish, $\mathscr{V}_{0} (x) = 0 , \mathscr{X}_{0} (x) = 0$.

In what follows, we adopt the polar coordinate system $(r , \theta , \varphi)$ for the spatial coordinates:
\begin{align}
g \mathscr{A}_{r} (x) = & 0 , \ \ \ 
g \mathscr{A}_{\theta} (x) = A (r) T_{\theta}, \ \ \ 
g \mathscr{A}_{\varphi} (x) = A (r) T_{\varphi} , \\
g \mathscr{V}_{r} (x) = & 0 , \ \ \ 
g \mathscr{V}_{\theta} (x) = V (r) T_{\theta} , \ \ \ 
g \mathscr{V}_{\varphi} (x) = V (r) T_{\varphi} , \\
g \mathscr{X}_{r} (x) = & 0 , \ \ \ 
g \mathscr{X}_{\theta} (x) = X (r) T_{\theta} , \ \ \  
g \mathscr{X}_{\varphi} (x) = X (r) T_{\varphi} 
,
\end{align}
where we have defined
\begin{align}
T_{\theta} = \frac{1}{2} \begin{pmatrix}
0 & i e^{- i \varphi} \\
- i e^{i \varphi} & 0 
\end{pmatrix}
 , \ \ \ \ 
T_{\varphi} = \frac{1}{2} \begin{pmatrix}
- \sin \theta & \cos \theta e^{- i \varphi} \\
\cos \theta e^{i \varphi} & \sin \theta
\end{pmatrix}
,
\end{align}
and 
\begin{equation}
A (r) = \frac{1 - f (M_{\mathscr{X}} r)}{r} , \ \ \ 
V (r) = \frac{1}{r} , \ \ \ 
X (r) = - \frac{f (M_{\mathscr{X}} r)}{r}
.
\end{equation}

\begin{figure}[t]
\centering
\includegraphics[width=0.45\textwidth]{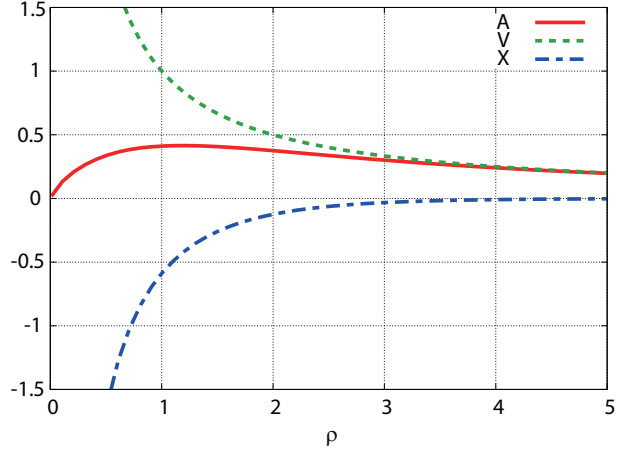}
\caption{The behaviors of $A ,V $ and $X $ as functions of $\rho = M_{\mathscr{X}} r$. 
Here $A(x)=V(x)+X(x)$ where $V (x)$ agrees with the Wu--Yang monopole
and $X (x)$ corresponds to the massive mode. 
}
\label{fields}
\end{figure}
 Fig.\ref{fields} is the plot of the fields $A , V $ and $X  $  as functions of $\rho = M_{\mathscr{X}} r$, which shows that the original gauge field $\mathscr{A} (x)$ is indeed regular at the origin
\begin{equation}
A (r) \approx 
M_{\mathscr{X}}^{2} \biggl[  - \widetilde{C} r - \frac{1}{3} r \log (M_{\mathscr{X}} r) + \mathcal{O} (r^{3} ) \biggr]
,
\end{equation}
as is expected.
On the other hand, the fields $\mathscr{V} (x)$ and $\mathscr{X} (x)$ diverge at the origin $r=0$.

We perform the singular gauge transformation which makes $\hat{\bm{\phi}} (x)$ diagonal: $\hat{\bm{\phi}}_{\infty} = \frac{1}{2} \sigma_{3}$
\begin{align}
\hat{\bm{\phi}} (x)  = \frac{1}{2} \begin{pmatrix}
\cos \theta & \sin \theta e^{- i \varphi} \\
\sin \theta e^{i \varphi} & - \cos \theta
\end{pmatrix} \to \hat{\bm{\phi}}^{\prime} (x) = U (x) \hat{\bm{\phi}} (x) U^{-1} (x) = \frac{1}{2} \begin{pmatrix}
1 & 0 \\
0 & -1
\end{pmatrix} =: \hat{\bm{\phi}}_{\infty}
,
\end{align}
or equivalently $\hat{\phi}^{\prime A} (x) = \delta^{A 3}$.
Such a gauge transformation can be done by using the following $SU(2)$ matrix $U(x)$:
\begin{align}
U (x) = \begin{pmatrix}
\cos \frac{\theta}{2} & \sin \frac{\theta}{2} e^{- i \varphi} \\
- \sin \frac{\theta}{2} e^{i \varphi} & \cos \frac{\theta}{2}
\end{pmatrix} \in SU (2)
.
\label{gauge_matrix}
\end{align}
As mentioned before, $\mathscr{X}_{\mu} (x)$ is transformed in an adjoint way:
\begin{equation}
\mathscr{X}_{\mu} (x) \to \mathscr{X}^{\prime}_{\mu} (x) = U (x) \mathscr{X}_{\mu} (x) U^{-1} (x)
,
\end{equation}
while, as a consequence, $\mathscr{V}_{\mu} (x)$ has the same gauge transformation property as the original gauge field $\mathscr{A}_{\mu} (x)$:
\begin{equation}
\mathscr{V}_{\mu} (x) \to \mathscr{V}_{\mu}^{\prime} = U (x) \left( \mathscr{V}_{\mu} (x) + \frac{i}{g} \partial_{\mu} \right) U^{-1} (x)
.
\end{equation}
Thus, $\mathscr{V} (x)$ and $\mathscr{X} (x)$ are transformed by $U (x)$ as
\begin{align}
g \mathscr{V}_{r}^{\prime} (x) = & 0 , \ \ \ 
g \mathscr{V}_{\theta}^{\prime} (x) = 0 , \ \ \ 
g \mathscr{V}_{\varphi}^{\prime} (x) = - \frac{1}{r} \frac{1 - \cos \theta}{\sin \theta} T_{3}, \\
g \mathscr{X}_{r}^{\prime} (x) = & 0 , \ \ \ 
g \mathscr{X}_{\theta}^{\prime} (x) = - \frac{f (M_{\mathscr{X}} r)}{r} T_{-} 
 , \ \ \  
g \mathscr{X}_{\varphi}^{\prime} (x) = - \frac{f (M_{\mathscr{X}} r)}{r} T_{+}
,
\end{align}
where we have defined
\begin{align}
T_{+} :=  T_{1} \cos \varphi + T_{2} \sin \varphi 
, \ \ \ 
T_{-} :=  T_{1} \sin \varphi - T_{2} \cos \varphi ,
 \quad T_A=\frac{\sigma_A}{2} 
.
\end{align}
One can find that the field $\mathscr{V} (x)$ is nothing but the Wu--Yang potential \cite{Wu--Yang} which has singularities of the Dirac string type \cite{Dirac} located on the negative part of the $z$-axis.
Moreover, we find that by recalling $f (M_{\mathscr{X}} r) \propto \exp (- M_{\mathscr{X}} r)$ at $r \approx \infty$ the field $\mathscr{X}(x)$ indeed falls off exponentially, and hence we can identify $\mathscr{X } (x)$ with the massive (or high energy) mode.

For the Yang--Mills magnetic monopole obtained in the massive Yang--Mills theory, we do not need to introduce the artificial regularization by hand to remedy the short-distance (or ultraviolet) singularity and instability of the Wu--Yang magnetic monopole in the pure massless Yang--Mills theory as worked out in \cite{regularize, Banks-Myerson-Kogut}. 
The regularized solution of the Yang--Mills field equation was obtained so that the Wu--Yang solution for  $r>r_0$  and another solution for $r < r_0$ are connected at $r = r_0$ to make the energy finite, see pp.503--504 and Appendix B of \cite{Banks-Myerson-Kogut}.
The Yang--Mills magnetic monopole $\mathscr{A}(x)$ obtained in this paper approaches the Wu--Yang type $\mathscr{V}(x)$ for large $r$, while for small $r$ it approaches the regular form and the energy becomes finite. 
This is attributed to the behavior of the massive mode $\mathscr{X}(x)$.  For large $r$, $\mathscr{X}(x)$ falls off quickly to guarantee $\mathscr{A}(x) \simeq \mathscr{V}(x)$, while for small $r$, $\mathscr{X}(x)$ also becomes singular but with the signature opposite to $\mathscr{V}(x)$ to cancel the singularity of $\mathscr{V}(x)$ leading to a finite  Yang-Mills field,   $\mathscr{A}(x)=\mathscr{V}(x)+\mathscr{X}(x) \simeq 0$ near $x=0$.

\subsection{The Chromo-Magnetic Field}

\begin{figure}[t]
\centering 
\includegraphics[width=0.45\textwidth]{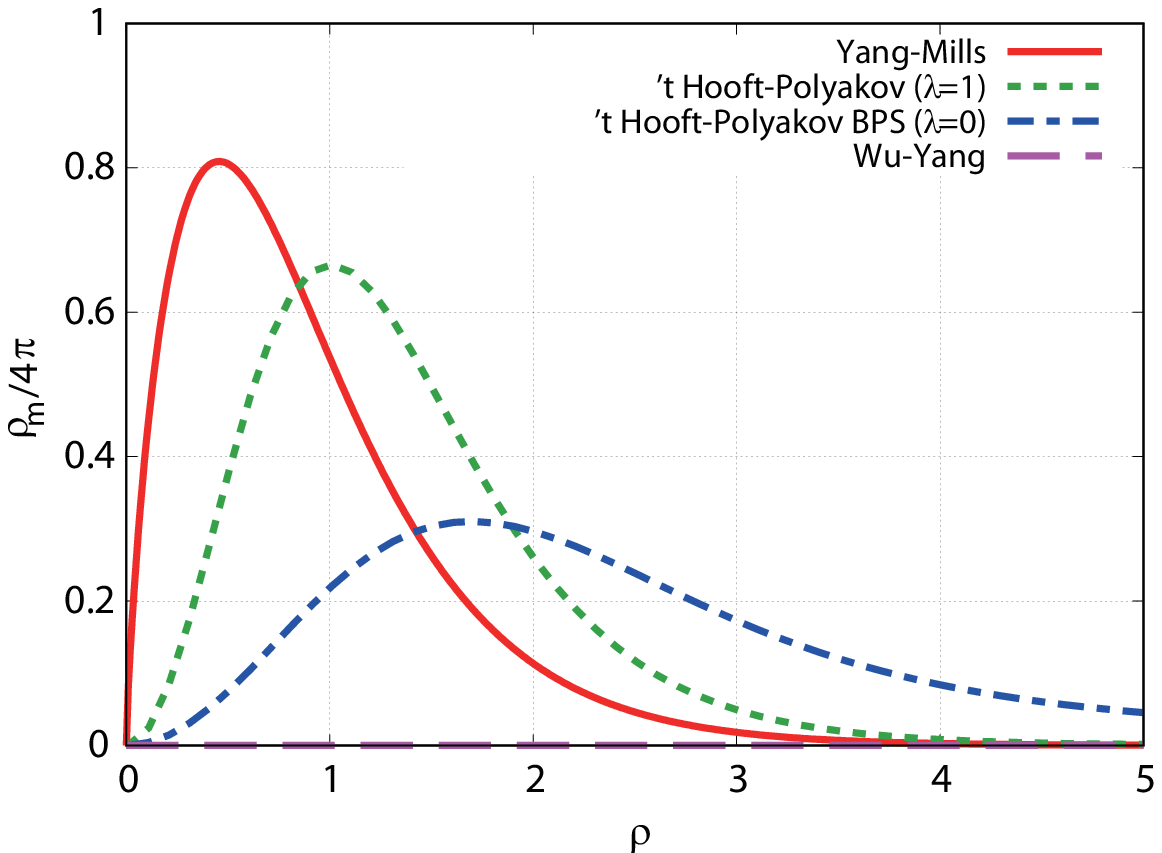} \ 
\includegraphics[width=0.45\textwidth]{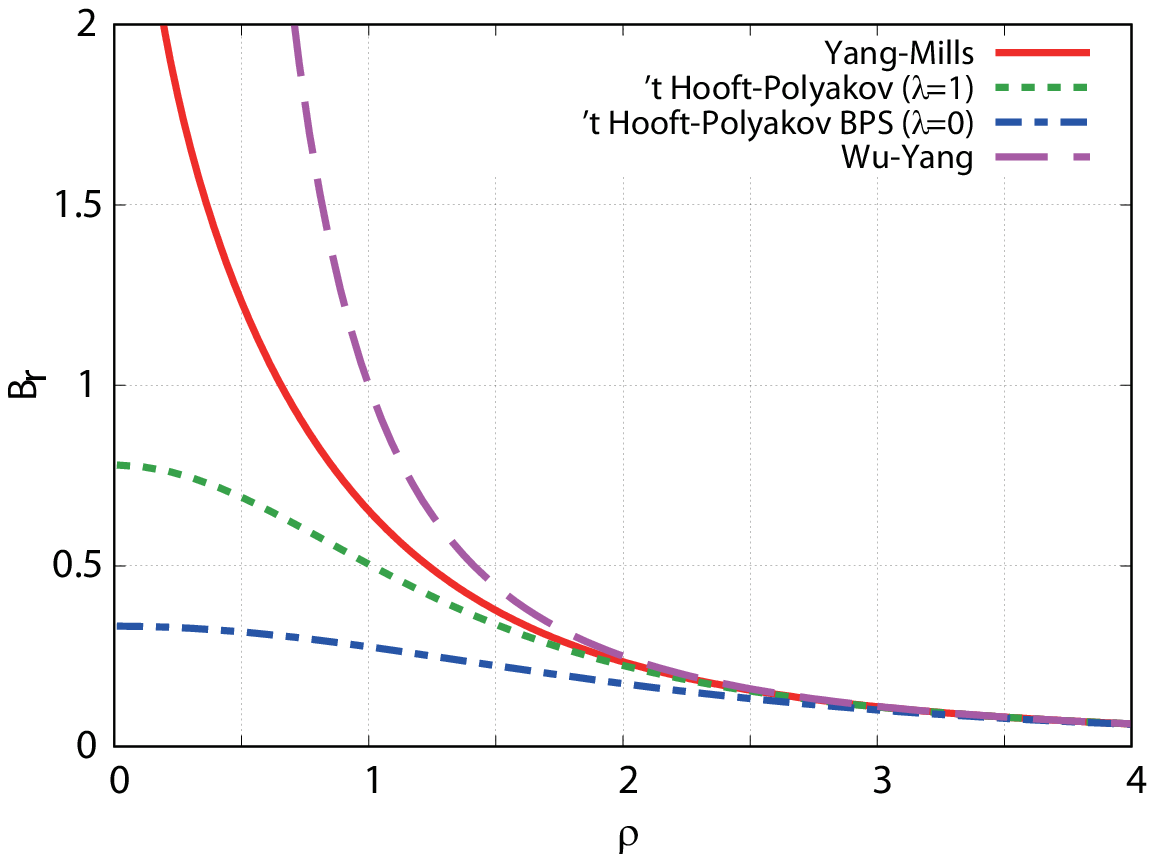} 
\caption{The short distance behaviors of (left) the magnetic charge density $\rho_{m}/ 4\pi$, (right) the gauge-invariant chromo-magnetic field $B_{r} := \mathscr{B}_{r}^{\prime} \cdot \hat{\bm{\phi}}_{\infty}$ as functions of $\rho = M_{\mathscr{X}} r$.
For the Wu--Yang monopole, the magnetic charge density is proportional to the delta function: $\rho_{m} \propto \delta (x)$.}
\label{magnetic_field}
\end{figure}

We examine the magnetic charge $q_{m}$ obtained by the chromo-magnetic field $\mathscr{B}_{j}^{A} (x)$:
\begin{equation}
g \mathscr{B}_{j}^{A} (x) = \frac{1}{2} \epsilon_{j k l} g \mathscr{F}_{k l}^{A} (x) =  \frac{x^{A} x^{j}}{r^{4}} \left( 1 - f^{2} (M_{\mathscr{X}} r)  \right) - \left( \frac{\delta^{A j}}{r} - \frac{x^{A} x^{j}}{r^{3}} \right) \frac{d f (M_{\mathscr{X}} r)}{d r}
.
\end{equation}
The magnetic charge $q_{m}$ and its density $\rho_{m} (r)$ are defined by
\begin{equation}
q_{m} = \int d^{3} x \ \mathscr{B}_{j}^{A} \left( \mathscr{D}_{j} [\mathscr{A}] \hat{\bm{\phi}} \right)^{A} = \int_{0}^{\infty} d r \ \rho_{m} (r)  , \ \ \ 
\rho_{m} (r) := 4 \pi  r^{2} \mathscr{B}_{j}^{A} \left( \mathscr{D}_{j} [\mathscr{A}] \hat{\bm{\phi}} \right)^{A}
.
\end{equation}
The magnetic charge density $\rho_{m} (r)$ can be written in terms of the profile functions $f (M_{\mathscr{X}} r)$ (and $h (M_{\mathscr{X}} r)$)
\begin{equation}
\rho_{m} (r) = \frac{4 \pi}{g} \frac{d}{d r} \bigl[ h (M_{\mathscr{X}} r) \left( 1 - f^{2} (M_{\mathscr{X}} r)  \right) \bigr]
.
\end{equation}
From the definition of $q_{m}$, this chromo-magnetic field $\mathscr{B}_{j}^{A} (x)$ indeed has a nontrivial magnetic charge $q_{m}$:
\begin{align}
q_{m} : = & \int_{0}^{\infty} d r \ \rho_{m} (r) = \frac{4 \pi}{g} \int_{0}^{\infty} d r \ \frac{d}{d r}  \bigl[ h (M_{\mathscr{X}} r) \left( 1 - f^{2} (M_{\mathscr{X}} r) \right) \bigr] 
= \frac{4 \pi}{g} \bigl[ h (M_{\mathscr{X}} r) \left( 1 - f^{2} (M_{\mathscr{X}} r) \right) \bigr] \biggl|_{r = 0}^{r = \infty} = \frac{4 \pi}{g}
.
\end{align}

In the left panel of Fig.\ref{magnetic_field} we give the magnetic charge density $\rho_{m} (r)$, which is also compared with the 't Hooft--Polyakov  magnetic monopole.
We observe that the Yang-Mills magnetic monopole is more localized in the vicinity of the origin than any 't Hooft--Polyakov  magnetic monopole.

In order to investigate the behavior of the chromo-magnetic field $\mathscr{B}_{j}^{A} (x)$ around $r \approx 0$, we turn to the polar coordinate representation:
\begin{align}
g \mathscr{B}_{r} (x) = & \frac{1 - f^{2} (M_{\mathscr{X}} r)}{r^{2}}  \frac{1}{2} \begin{pmatrix}
\cos \theta & \sin \theta e^{- i \varphi} \\
\sin \theta e^{i \varphi} & - \cos \theta
\end{pmatrix}
, 
\nonumber\\
g \mathscr{B}_{\theta} (x) = & - \frac{1}{r} \frac{d f (M_{\mathscr{X}} r)}{d r}   \frac{1}{2} \begin{pmatrix}
\sin \theta & - \cos \theta e^{- i \varphi} \\
- \cos \theta e^{i \varphi} & - \sin \theta
\end{pmatrix}
, \ \ \ 
g \mathscr{B}_{\varphi} (x) = - \frac{1}{r} \frac{d f (M_{\mathscr{X}} r)}{d r}  \frac{1}{2} \begin{pmatrix}
0 & i e^{- i \varphi} \\
- i e^{i \varphi} & 0 
\end{pmatrix}
.
\end{align}

Then, $\mathscr{B} (x)$ is transformed by $U(x)$ in (\ref{gauge_matrix}), $\mathscr{B} (x) \to \mathscr{B}^{\prime} (x) = U (x) \mathscr{B} (x) U^{-1} (x)$:
\begin{align}
g \mathscr{B}_{r}^{\prime} (x) =  \frac{1 - f^{2} (M_{\mathscr{X}} r) }{r^{2}} T_{3} , 
\ \ \ 
g \mathscr{B}_{\theta}^{\prime} (x) =  - \frac{1}{r} \frac{d f (M_{\mathscr{X}} r)}{d r}  T_{+} , \ \ \ 
g \mathscr{B}_{\varphi}^{\prime} (x) = \frac{1}{r} \frac{d f (M_{\mathscr{X}} r)}{d r}  T_{-}
.
\end{align}
For $\mathscr{B} (x)$ to be gauge invariant, we take the inner product $\mathscr{B}^{\prime} \cdot \hat{\bm{\phi}}_{\infty}$, 
\begin{align}
g \mathscr{B}_{r}^{\prime} (x) \cdot \hat{\bm{\phi}}_{\infty} = \frac{1 - f^{2} (M_{\mathscr{X}} r)}{r^{2}} , \ \ \ 
g \mathscr{B}_{\theta}^{\prime} (x) \cdot \hat{\bm{\phi}}_{\infty} = g \mathscr{B}_{\varphi}^{\prime} (x) \cdot \hat{\bm{\phi}}_{\infty} = 0
.
\end{align}

We find that in the radially fixed case of Yang--Mills theory, the chromo-magnetic field diverges at the origin due to its logarithmic behavior:
\begin{equation}
g \mathscr{B}_{r}^{\prime} (x) \cdot \hat{\bm{\phi}}_{\infty} = \frac{1 - f^{2} (M_{\mathscr{X}} r)}{r^{2}} \approx M_{\mathscr{X}}^{2} \biggl[ - \frac{2}{3} \log (M_{\mathscr{X}} r)  + ({\rm finite \ terms}) \biggr]
.
\end{equation}
See the right panel of Fig.\ref{magnetic_field}.
This magnetic field $\mathscr{B}_{r}^{\prime} (x) \cdot \hat{\bm{\phi}}_{\infty}$ should be compared with that of the 't Hooft--Polyakov monopole which has a finite value  even at the origin. 

It should be noticed that we have included the volume element $4 \pi r^{2}$ in the definition of the energy and magnetic charge densities because we started from the Lagrangian including $4 \pi r^{2}$.
By excluding the factor $r^{2}$ from the energy and magnetic charge densities, in the radially fixed case, they diverge at the origin $r = 0$ due to the logarithmic term $\log (M_{\mathscr{X}} r)$.
This divergence, however, is not essential for the calculation of the physical quantities, since, for example, in order to evaluate the magnetic charge $q_{m}$ we need the volume element $4 \pi r^{2}$, which makes the divergence disappear.

\section{conclusion and discussion}

In this paper we have constructed the magnetic monopole {\it configuration} in the $SU(2)$ Yang--Mills theory even in the absence of the scalar field by incorporating a gauge-invariant mass term. 
Such a gauge-invariant mass term is obtained through a gauge-independent description of the BEH mechanism proposed in \cite{Kondo2016}. 
The procedure for obtaining the relevant magnetic monopole is guided by the ``complementarity'' between the $SU(2)$ gauge-adjoint scalar model with the radial-fixing constraint and the massive $SU(2)$ Yang--Mills theory \cite{Kondo2016}.
In fact, we have obtained the static and spherically symmetric magnetic monopole configuration in the $SU(2)$ massive Yang--Mills theory by solving the field equations of the ``complementary'' $SU(2)$ gauge-adjoint scalar model with a radially fixed scalar field.
We have found that the static energy or the rest mass of the obtained Yang--Mills magnetic monopole is finite and proportional to the mass $M_{\mathscr{X}}$ of the massive components $\mathscr{X}$ of the Yang--Mills gauge field $\mathscr{A}$.

In the long-distance region, we observed that the Yang--Mills magnetic monopole configuration $\mathscr{A}$ reduces to the restricted field $\mathscr{V}$ which agrees with the Wu--Yang magnetic monopole as a consequence of the suppression of the massive modes $\mathscr{X}$ in the long-distance region.
This feature is similar to the usual 't Hooft--Polyakov monopoles.
In the short-distance region, on the other hand, the Wu--Yang magnetic monopole becomes singular, while  the 't Hooft--Polyakov monopole remains non-singular even at the origin. 
In the Yang--Mills magnetic monopole, we found that the massive components $\mathscr{X}$ play the very important role for canceling the singularity of $\mathscr{V}$ in the short-distance region such that the original gauge field $\mathscr{A}$ remains non-singular at the origin.
This regularity of the Yang--Mills magnetic monopole is guaranteed by the logarithmic behavior of the gauge field itself without an aid of the scalar field which vanishes at the origin as seen in 't Hooft--Polyakov monopoles. 
This behavior renders the energy of the Yang--Mills magnetic monopole finite even if the magnitude of the scalar field is fixed. 
It should be remarked that the chromo-magnetic field $\mathscr{B}$ is divergent at the origin due to the logarithmic behavior of the solution $f (\rho)$, which is however unessential for obtaining finite physical quantities such as energy, magnetic charge density and magnetic flux.

By using the Yang--Mills magnetic monopole found in this paper, we can show quark confinement in the three-dimensional Yang--Mills theory in the same way as the three-dimensional Georgi--Glashow model shown by Polyakov \cite{Polyakov} without introducing the artificially regularized Yang--Mills magnetic monopole \cite{regularize,Banks-Myerson-Kogut} for avoiding the short-distance singularity and instability of the Wu--Yang magnetic monopole.

\section*{Acknowledgments}

S.N. would like to thank Nakamura Sekizen-kai for a scholarship. R. M. was  supported by Grant-in-Aid for JSPS Research Fellow Grant Number 17J04780. M.W. was supported by the Ministry of Education, Culture, Sports, Science and Technology, Japan (MEXT scholarship). K.-I. K. was  supported by Grant-in-Aid for Scientific Research, JSPS KAKENHI Grant Number (C) No.15K05042.

\appendix
\section{The Georgi--Glashow model and the 't Hooft--Polyakov monopole: radially variable case}
\label{AppendixA}
We have used the Georgi--Glashow model to compare the radially fixed monopole with the radially variable one.
We introduce the Georgi--Glashow model by the Lagrangian density:
\begin{equation}
\mathscr{L}_{\rm GG} =  - \frac{1}{4} \mathscr{F}_{\mu \nu} \cdot \mathscr{F}^{\mu \nu} + \frac{1}{2} \left( \mathscr{D}_{\mu} [\mathscr{A}] \bm{\phi} \right) \cdot \left( \mathscr{D}^{\mu} [\mathscr{A}] \bm{\phi} \right) - \frac{\lambda^{2} g^{2}}{4} \left( \bm{\phi} \cdot \bm{\phi} - v^{2} \right)^{2} 
.
\end{equation}
We take the usual ansatz for the 't Hooft--Polyakov monopole:
\begin{equation}
g \mathscr{A}_{0}^{A} (x) = 0 , \ \ \ 
g \mathscr{A}_{j}^{A} (x) = \frac{\epsilon^{j A k} x^{k}}{r^{2}} ( 1 - f (r) ) , \ \ \ 
\phi^{A} (x) = v \frac{x^{A}}{r} h (r)
.
\end{equation}

By redefining $\widetilde{\mathscr{L}}_{\rm GG} = 4 \pi r^{2} \mathscr{L}_{\rm GG}$ and rescaling the variable $r \to \rho := g v r$ to be dimensionless, we have
\begin{equation}
\widetilde{\mathscr{L}}_{\rm GG} = 4\pi v^{2} \biggl[ - f^{\prime 2} (\rho) - \frac{ ( f^{2} (\rho) - 1 )^{2}}{2 \rho^{2}} - \frac{1}{2} \rho^{2} h^{\prime 2} (\rho) - f^{2} (\rho) h^{2} (\rho) - \frac{\lambda^{2}}{4} \rho^{2} \left( h^{2} (\rho) - 1 \right)^{2} \biggr]
.
\end{equation}
The field equations are obtained as
\begin{align}
f^{\prime \prime} (\rho) = & \frac{f^{3} (\rho) - f (\rho)}{\rho^{2}} + f (\rho) h^{2} (\rho) ,
\label{eq_GG_f} \\
\left( \rho^{2} h^{\prime} (\rho) \right)^{\prime} = & 2 f^{2} (\rho) h (\rho) + \lambda^{2} \rho^{2} \left( h^{3} (\rho) - h (\rho) \right)
\label{eq_GG_h}
.
\end{align}
We assume the asymptotic behavior for small $\rho$ as the power series in $\rho$:
\begin{align}
f (\rho) \approx 1 + \sum_{n = 0}^{\infty} F_{n} \rho^{n} , \ \ \ \ 
h (\rho) \approx \sum_{n=0}^{\infty} H_{n} \rho^{n}
.
\end{align}
By substituting these series expansions into the field equations (\ref{eq_GG_f}) and (\ref{eq_GG_h}), we can determine the coefficients as
\begin{align}
f (\rho) \approx & 1 - F_{2} \rho^{2} + \frac{3 F_{2}^{2} + H_{1}^{2} }{10} \rho^{4} - \frac{14 F_{2}^{3} + 12 F_{2} H_{1}^{2} + \lambda H_{1}^{2}}{140} \rho^{6} + \cdots, \\
h (\rho) \approx & H_{1} \rho - \frac{\lambda H_{1} + 4 F_{2} H_{1}}{10} \rho^{3} + \frac{48 F_{2}^{2} H_{1} + 8 \lambda F_{2} H_{1} + \lambda^{2} H_{1} + 2 ( 5 \lambda + 2 ) H_{1}^{3}}{280} \rho^{5} + \cdots
.
\end{align}
Notice that the logarithmic behavior does not appear since the field equations are satisfied by the power series in $\rho$ without the logarithmic terms for $0 \leq \lambda < \infty$ in the case of the 't Hooft--Polyakov monopoles.

\section{A numerical treatment}
\label{AppendixB}
We have used the integral equation method \cite{Bais-Primack} to solve the monopole equation.
We can rewrite the monopole equation (\ref{eq_f3}) as
\begin{equation}
\mathcal{L}_{\rho} f (\rho) = f^{3} (\rho) - f (\rho) + \rho^{2} f (\rho) + 2 \rho f^{\prime} (\rho) - ( \rho^{2} + \nu ( \nu + 1 ) ) f (\rho)
,
\end{equation}
where we have introduced the differential operator $\mathcal{L}_{\rho}$ defined by
\begin{equation}
\mathcal{L}_{\rho} := \rho^{2} \frac{d^{2}}{d \rho^{2}} + 2 \rho \frac{d}{d \rho} - ( \rho^{2} + \nu ( \nu + 1 ))
.
\end{equation}
By using the Green's function $G (\rho ,s)$ for the differential operator $\mathcal{L}_{\rho}$, i.e., $\mathcal{L}_{\rho} G (\rho ,s) = - \delta (\rho - s)$, which is given using the modified spherical Bessel functions as 
\begin{align}
G (\rho , s) = \left\{ \begin{array}{cc}
k_{\nu} (\rho) i_{\nu} (s) & (s < \rho) \\
k_{\nu} (s) i_{\nu} (\rho) & (\rho > s) 
\end{array} \right.
,
\end{align}
the equation (\ref{eq_f3}) can be rewritten into the integral equation
\begin{equation}
f (\rho) = \int_{0}^{\infty} d s \ G (\rho, s) \mathfrak{F} (f (s) , f^{\prime} (s) , s)
,
\label{int_eq}
\end{equation}
where $\mathfrak{F}$ stands for the inhomogeneous terms
\begin{equation}
\mathfrak{F} ( f (s) , f^{\prime} (s) , s) = f^{3} (s) - f (s) + s^{2} f (s) + 2 s f^{\prime} (s) - ( s^{2} + \nu ( \nu + 1 ) ) f (s)
.
\end{equation}
Here we have chosen the modified spherical Bessel functions for accelerating the convergence of numerical calculations.

According to \cite{Bais-Primack}, we solve the integral equation (\ref{int_eq}) by numerical iterations:
\begin{equation}
f^{(n)} (\rho) = f^{(0)} (\rho) + \int_{0}^{\infty} d s \ G( \rho, s) \left\{ \mathfrak{F} ( f (s) , f^{\prime} (s) , s) + \mathcal{L}_{s} f^{(0)} (s) \right\}
,
\end{equation}
where $f^{(0)} (\rho)$ is a trial function for numerical calculations and $f^{(n)} (\rho)$ is the appropriate solution obtained at $n$-iterations. 
We have chosen $f^{(0)} (\rho)$ to satisfy the boundary conditions of the original $f (\rho)$:
\begin{equation}
f^{(0)} (\rho) = \mathrm{sech} \ \rho
.
\end{equation}
After $n = 30$ iterations, we find that $f^{(n)} (\rho)$ converges to a reliable result.
For the 't Hooft--Polyakov monopole in the Georgi--Glashow model, we can perform the same procedure as that given in \cite{Bais-Primack}.
The result of our numerical calculations is consistent with the 't Hooft--Polyakov monopole \cite{numerical,Vinciarelli} for the infinite coupling $\lambda=\infty$ in the Georgi--Glashow model.

\end{document}